\magnification \magstep1
\raggedbottom
\openup 1\jot
\voffset6truemm
\headline={\ifnum\pageno=1\hfill\else
\hfill {\it Foundational problems in quantum gravity} 
\hfill \fi}
\centerline {\bf FOUNDATIONAL PROBLEMS IN QUANTUM GRAVITY}
\vskip 1cm
\leftline {Ivan G. Avramidi$^{1}$ and Giampiero Esposito$^{2}$}
\vskip 0.3cm
\noindent
${ }^{1}${\it Department of Mathematics, The University of Iowa,
14 MacLean Hall, Iowa City, IA 52242, USA}
\vskip 0.3cm
\noindent
${ }^{2}${\it Istituto Nazionale di Fisica Nucleare, Sezione di Napoli,
Mostra d'Oltremare Padiglione 20, 80125 Napoli, Italy}
\vskip 1cm
\noindent
Boundary conditions play a crucial role in the path-integral
approach to quantum gravity and quantum cosmology, as well as
in the current attempts to understand the one-loop semiclassical
properties of quantum field theories. Within this framework,
one is led to consider boundary conditions completely invariant
under infinitesimal diffeomorphisms on metric perturbations. These
are part of a general scheme, which can be developed for Maxwell
theory, Yang--Mills Theory, Rarita--Schwinger fields and any
other gauge theory. A general condition for strong ellipticity
of the resulting field theory on manifolds with boundary is 
here proved, following recent work by the authors. The relevance
for Euclidean quantum gravity is eventually discussed.
\vskip 6cm
\noindent
Contribution to the Italian XIII National Conference on General
Relativity, Monopoli, September 1998.
\vskip 100cm
\noindent
In several branches of classical and quantum field theory, as well
as in the current attempts to develop a quantum theory of the
universe and of gravitational interactions, it remains very useful
to describe physical phenomena in terms of differential equations
for the variables of the theory, supplemented by boundary 
conditions for the solutions of such equations. For example, the
problems of electrostatics, the analysis of waveguides, the theory
of vibrating membranes, the Casimir effect, van der Waals forces,
and the problem of how the universe could evolve from an initial
state, all need a careful assignment of boundary conditions. In
the latter case, if one follows a path-integral approach, one faces
two formidable tasks.
\vskip 0.3cm
\noindent
(i) the classification of the geometries occurring in the ``sum
over histories'' and matching the assigned boundary data;
\vskip 0.3cm
\noindent
(ii) the choice of boundary conditions on metric perturbations 
which may lead to the evaluation of the one-loop semiclassical
approximation. Indeed, while the full path integral for quantum
gravity is a fascinating idea but remains a formal tool, the
one-loop calculation may be put on solid ground, and appears
particularly interesting because it yields the first quantum
corrections to the underlying classical theory (despite the
well known lack of perturbative renormalizability of quantum
gravity based on Einstein's theory). Within this framework, it
is of crucial importance to understand whether the property of
strong ellipticity of the boundary-value problem 
(see Appendix) is compatible with the 
request of local and gauge-invariant boundary conditions
for a self-adjoint operator on perturbations.
\vskip 0.3cm
For this purpose, we are now going to study gauge-invariant 
boundary conditions in a general gauge theory, following
Ref. [1]. Given a Riemannian manifold $M$, a
gauge theory is defined by two vector bundles, 
$V$ and $G$, such that ${\rm dim}\,V> {\rm dim}\,G$.
$V$ is the bundle of gauge fields $\varphi\in C^\infty(V,M)$, 
and $G$ is the bundle 
of parameters of gauge transformations $\epsilon\in C^\infty(G,M)$.
Both bundles $V$ and $G$ are equipped with some Hermitian 
positive-definite metrics $E$, $E^{\dag}=E$, and $\gamma$, 
$\gamma^{\dag}=\gamma$,
and with the corresponding natural $L^2$ scalar products
$(,)_V$ and $(,)_G$.

The infinitesimal gauge transformations
$$
\delta \varphi=R\epsilon
\eqno (1)
$$
are generated by a first-order differential operator $R$,
$$
R:\ C^\infty(G,M)\to C^\infty(V,M).
\eqno (2)
$$
Furthermore, two auxiliary operators are introduced,
$$
{X}:\ C^\infty(V,M)\to C^\infty(G,M)
\eqno (3)
$$
and
$$
{Y}:\ C^\infty(G,M)\to C^\infty(G,M),
\eqno (4)
$$
which make it possible to define the differential operators
$$
{L} \equiv {X} R:\ C^\infty(G,M)\to C^\infty(G,M) 
\eqno (5)
$$
and
$$
H \equiv \bar{X}{Y}{X}:\ C^\infty(V,M)\to C^\infty(V,M),
\eqno (6)
$$
where $\bar{X} \equiv E^{-1}{X}^{\dag}\gamma$.
The operators $X$ and $Y$ should satisfy the following conditions
(but are otherwise arbitrary):
\vskip 0.3cm
(1) The differential operators $L$ and $H$ have the same order.
\vskip 0.3cm
(2) The operators $L$ and $H$ are formally 
self-adjoint (or anti-self-adjoint).
\vskip 0.3cm
(3) The operators $L$ and $Y$ are elliptic.
\vskip 0.3cm
{}From these conditions, two essentially different cases are found:
\vskip 0.3cm
\noindent
{\bf Case I}. $X$ is of first order and $Y$ is of zeroth order, i.e.
$$
X=\bar R, \; \; \; \; Y=I_{G},
\eqno (7)
$$
where $\bar R \equiv \gamma^{-1}R^{\dag}E$. Then, of course,
$L$ and $H$ are both {\it second-order} differential operators,
$$
L=\bar R R, \; \; \; \; H=R\bar R.
\eqno (8)
$$
\vskip 0.3cm
\noindent
{\bf Case II}. ${X}$ is of zeroth order and ${Y}$ is of first order.
Let ${\cal R}$ be the bundle of maps of $G$ into $V$, and let
$\beta\in {\cal R}$ be a zeroth-order differential operator. Then
$$
X=\bar \beta, \; \; \; \; Y=\bar\beta R ,
\eqno (9)
$$
where $\bar\beta \equiv \gamma^{-1}\beta^{\dag}E$,
and the operators $L$ and $H$ are of {\it first} order,
$$
L=\bar\beta R, \; \; \; \; H=\beta\bar\beta R\bar\beta=\beta L\bar\beta.
\eqno (10)
$$
We assume that, by suitable choice of the parameters,
the second-order operator $\bar R R$ can be made
of Laplace type and the first-order operator 
$\bar\beta R$ can be made of Dirac type,
and, therefore, have non-degenerate leading symbols (here denoted
by $\sigma_{L}$) 
$$
{\rm det}_G\sigma_L(\bar R R)\ne 0,
\eqno (11)
$$
$$
{\rm det}_G\sigma_L(\bar \beta R)\ne 0.
\eqno (12)
$$
The dynamics of gauge fields $\varphi\in C^\infty(V,M)$ 
at the linearized (one-loop) level is described by a
formally self-adjoint (or anti-self-adjoint) differential operator,
$$
\Delta:\ C^\infty(V,M)\to C^\infty(V,M).
\eqno (13)
$$
This operator is of second order for bosonic fields and of 
first order for fermionic fields.
In both cases its leading symbol satisfies the identities
$$
\Delta R = 0, \; \; \; \; 
\bar R \Delta=0,
\eqno (14)
$$
and, therefore, is degenerate.

We consider only the case when the gauge generators are
{\it linearly independent}. This means that the equation
$$
\sigma_L(R)\epsilon=0,
\eqno (15)
$$
has {\it the only} solution $\epsilon=0$. 
In other words, 
$$
{\rm Ker} \; \sigma_L(R)=\emptyset,
\eqno (16)
$$
i.e. the rank
of the leading symbol of the operator $R$ equals the dimension
of the bundle $G$,
$$
{\rm rank}\, \sigma_L(R)=\dim\,G.
\eqno (17)
$$
We also assume that the leading symbols of the generators $R$ 
are {\it complete} in that they generate {\it all} 
zero-modes of the leading
symbol of the operator $\Delta$, i.e. {\it all} solutions of the equation
$$
\sigma_L(\Delta)\varphi=0,
\eqno (18)
$$
have the form
$$
\varphi=\sigma_L(R)\epsilon,
\eqno (19)
$$
for some $\epsilon$. In other words,
$$
{\rm Ker} \; \sigma_L(\Delta)
=\{\sigma_L(R)\epsilon\ |\ \epsilon\in G\},
\eqno (20)
$$
and hence
$$
{\rm rank} \; \sigma_L(\Delta)={\rm dim}\,V
-{\rm dim}\,G.
\eqno (21)
$$
Furthermore, let us take the operator $H$ of the 
{\it same} order as the operator $\Delta$
and construct a formally (anti-)self-adjoint operator, 
$$
F \equiv \Delta+H,
\eqno (22)
$$
so that
$$
\sigma_{L}(F)=\sigma_{L}(\Delta)+\sigma_{L}(H).
\eqno (23)
$$

It is easy to derive the following result: 
\vskip 0.3cm
\noindent
The leading symbol of the operator $F$ is non-degenerate, i.e.
$$
{\rm det}_V\sigma_L(F)\ne 0.
\eqno (24)
$$
\vskip 0.3cm
\noindent
{\bf Proof}. Indeed, suppose there exists a zero-mode
$\varphi_0$ of the leading symbol of the operator $F$, i.e.
$$
\sigma_L(F)\varphi_0=\bar\varphi_0\sigma_L(F)=0,
\eqno (25)
$$
where
$\bar\varphi\equiv \varphi^{\dag}E$.
Then we have
$$
\bar\varphi_0\sigma_L(F)\sigma_L(R)
=\bar\varphi_0\sigma_L(\bar XY)\sigma_L({L})=0,
\eqno (26)
$$
and, since $\sigma_L({L})$ is non-degenerate,
$$
\bar\varphi_0\sigma_L(\bar XY)=\sigma_L(YX)\varphi_0=0.
\eqno (27)
$$
But this implies
$$
\sigma_L(H)\varphi_0=0,
\eqno (28)
$$
and hence
$$
\sigma_L(F)\varphi_0=\sigma_L(\Delta)\varphi_0=0.
\eqno (29)
$$
Thus, $\varphi_0$ is a zero-mode of the leading 
symbol of the operator $\Delta$,
and according to the completeness of the generators $R$ must have the form
$\varphi_0=\sigma_L(R)\epsilon$ for some $\epsilon$.
Substituting this form into Eq. (27) we obtain
$$
\sigma_L(YX)\sigma_L(R)\epsilon
=\sigma_L({YL})\epsilon=0.
\eqno (30)
$$
Herefrom, by taking into account the non-singularity of $\sigma_L({YL})$,
it follows $\epsilon=\varphi_0=0$, and hence the leading symbol of the 
operator $F$ has no zero-modes, i.e. it is non-degenerate.

Thus, the operators ${L}$ and $F$ have, 
both, non-degenerate leading symbols.
In quantum field theory the operator $X$ is called the gauge-fixing operator, 
$F$ the gauge-field operator, the operator ${L}$ 
the (Faddeev--Popov) ghost operator
and the operator $Y$ in the Case II the third 
(or Nielsen--Kallosh) ghost operator.
The most convenient and the most important case is when,
by suitable choice of the parameters it turns out to be possible to
make both the operators $F$ and ${L}$ either of Laplace type
or of Dirac type. The one-loop effective action for  
gauge fields is given by the functional
superdeterminants of the gauge-field operator $F$ 
and the ghost operators $L$ and $Y$: 
$$
\Gamma={1\over 2}\log({\rm Sdet}\, F)
-\log({\rm Sdet}\,L)
-{1\over 2}\log({\rm Sdet}\,Y).
\eqno (31)
$$

Let us now focus on bosonic fields, when $\Delta$ is a 
second-order formally self-adjoint operator. 
The gauge invariance identity (14) means, in particular,
$$
\sigma_L(\Delta)\sigma_L(R) = 0.
\eqno (32)
$$
Now we assume that both the operators $L=\bar RR$ and $F=\Delta+R\bar R$ 
are of Laplace type, i.e.
$$
\sigma_L(\bar R R)=|\xi|^{2} I_{G},
\eqno (33)
$$
$$
\sigma_L(F)=\sigma_L(\Delta)+\sigma_L(R\bar R)=|\xi|^{2} I_V .
\eqno (34)
$$
On manifolds with boundary one has to impose some boundary conditions 
to make these operators self-adjoint and elliptic. They read
$$
B_L\psi(\epsilon)=0,
\eqno (35)
$$
$$
B_F\psi(\varphi)=0,
\eqno (36)
$$
where $\psi(\epsilon)$ and $\psi(\varphi)$ are the 
boundary data (see Appendix) for the bundles
$G$ and $V$, respectively, and $B_{L}$ and $B_F$ are the corresponding
boundary operators. In gauge theories one tries to choose the 
boundary operators $B_{L}$ and $B_F$
in a gauge-invariant way, so that the condition
$$
B_F\psi(R\epsilon) = 0
\eqno (37)
$$
is satisfied identically 
for any $\epsilon$ subject to the boundary conditions (35).
This means that the boundary operators $B_L$ and $B_F$ satisfy the identity
$$
B_F [\psi,R](I_G-B_{L})\equiv 0,
\eqno (38)
$$
where $[\psi,R] \equiv \psi R - R \psi$.

We will see that this requirement fixes completely the form of the
as yet unknown boundary operator $B_{L}$. Indeed,
the most natural way to satisfy the condition 
of gauge invariance is as follows.
Let us decompose the cotangent bundle $T^{*}(M)$ in
such a way that $\xi=(N,\zeta)\in T^{*}(M)$, where 
$N$ is the inward-pointing unit normal to the boundary and
$\zeta\in T^{*}(\partial M)$ is a cotangent vector on the boundary.
Consider the restriction $W_0$ of the vector bundle $V$ to the boundary.
Let us define restrictions of the leading symbols 
of the operators $R$ and $\Delta$ to the boundary, i.e.
$$
\Pi\equiv \sigma_L(\Delta;N)\Big|_{\partial M} ,
\eqno (39)
$$
$$
\nu\equiv \sigma_L(R;N)\Big|_{\partial M},
\eqno (40)
$$
$$
\mu\equiv \sigma_L(R;\zeta)\Big|_{\partial M}.
\eqno (41)
$$
{}From Eq. (32) we have thus the identity
$$
\Pi\nu=0,
\eqno (42)
$$
Moreover, from (33) and (34) we have also
$$
\bar\nu\nu=I_{G} ,
\eqno (43)
$$
$$
\bar\nu\mu+\bar\mu\nu=0,
\eqno (44)
$$
$$
\bar\mu\mu=|\zeta|^{2} I_{G} ,
\eqno (45)
$$
$$
\Pi=I_V -\nu\bar\nu .
\eqno (46)
$$
{}From (42) and (43)
we find that $\Pi:\ W_0\to W_0$ is a self-adjoint 
projector orthogonal to $\nu$,
$$
\Pi^2=\Pi, \; \; \; \; \Pi\nu=0, \; \; \; \; \bar\Pi=\Pi.
\eqno (47)
$$
Then, a part of the boundary conditions for the operator $F$ reads
$$
\Pi\varphi\Big|_{\partial M}=0.
\eqno (48)
$$
The gauge transformation of this equation is
$$
\Pi R\epsilon\Big|_{\partial M}=0.
\eqno (49)
$$
The normal derivative does not contribute to
this equation, and hence, if 
Dirichlet boundary conditions are imposed on $\epsilon$,
$$
\epsilon\Big|_{\partial M}=0,
\eqno (50) 
$$
the equation (49) is identically satisfied.

The easiest way to get the other part of the 
boundary conditions is just to set
$$
\bar R\varphi\Big|_{\partial M}=0.
\eqno (51)
$$
Bearing in mind eq. (5) we find that, 
under the gauge transformations (1), 
this is transformed into
$$
{L}\epsilon\Big|_{\partial M}=0.
\eqno (52)
$$
If some $\epsilon$ is a zero-mode of the operator 
${L}$, i.e. $\epsilon\in {\rm Ker}\,{L}$, 
this is identically zero. For all $\epsilon\notin {\rm Ker}\,{L}$ 
this identically vanishes when the Dirichlet boundary conditions (50)
are imposed. In other words, the requirement of gauge
invariance of the boundary conditions 
(36) determines in an almost unique way (up to zero-modes)
that the ghost boundary operator $B_L$ should be of Dirichlet type. 
Anyway, Dirichlet boundary conditions
for the operator $L$ are {\it sufficient} to achieve
gauge invariance of the boundary conditions for the operator $F$.

Since the operator ${\bar R}$ in the boundary conditions 
(51) is a first-order
operator, the set of boundary conditions (48) and (51)
follows the general scheme formulated in Ref. [1].
Separating the normal derivative in the operator 
${\bar R}$ we find exactly the Gilkey--Smith boundary conditions
[2] for operators of Laplace type:
$$
\pmatrix{\Pi & 0 \cr \Lambda & I_V - \Pi \cr}
\pmatrix{[\varphi]_{\partial M} \cr 
[\varphi_{;N}]_{\partial M} \cr}=0,
\eqno (53)
$$
where
$$
\Lambda \equiv (I_V-\Pi) \left[{1\over 2}\Bigr(\Gamma^{i}
{\widehat \nabla}_{i}+{\widehat \nabla}_{i}\Gamma^{i}\Bigr)
+S \right](I_V-\Pi),
\eqno (54)
$$
the matrices $\Gamma^i$ having the form
$$
\Gamma^{i}=-\nu\bar\nu\mu^{i}\bar\nu,
\eqno (55)
$$
and $S$ being an endomorphism. The matrices
$\Gamma^{i}$ are anti-self-adjoint, $\bar\Gamma^i=-\Gamma^i$, 
and satisfy the relations
$$
\Pi\Gamma^i=\Gamma^i\Pi=0.
\eqno (56)
$$
Thus, one can now define the matrix 
$$
T \equiv \Gamma\cdot\zeta 
=-\nu\bar\nu\mu\bar\nu,
\eqno (57)
$$
where $\mu \equiv \mu^j\zeta_j$, and 
study the condition of strong ellipticity $|\zeta|I_V -iT >0$
obtained in Ref. [1]. Such a condition now reads
$$
|\zeta| I_V-iT=|\zeta| I_V +i\nu\bar\nu\mu\bar\nu>0.
\eqno (58)
$$
 
Moreover, using the eqs. (55), (44) and (45) we evaluate
$$
\Gamma^{(i}\Gamma^{j)}
=-(I_V-\Pi)\mu^{(i}\bar\mu^{j)}(I_V-\Pi).
\eqno (59)
$$
Therefore
$$
{T}^2=\Gamma^i\Gamma^j\zeta_i\zeta_j=
-(I_V-\Pi)\mu\bar\mu(I_V-\Pi),
\eqno (60)
$$
and
$$
T^2+|\zeta|^{2} I_V=|\zeta|^{2}\Pi
+(I_V-\Pi)[|\zeta|^{2} I_V
-\mu\bar\mu](I_V-\Pi).
\eqno (61)
$$
Since for non-vanishing $\zeta$ the part proportional 
to $\Pi$ is positive-definite,
the condition of strong ellipticity for bosonic gauge theory means
$$
(I_V-\Pi)[|\zeta|^{2} I-\mu\bar\mu](I_V-\Pi)>0.
\eqno (62)
$$

We have thus proved a theorem:
\vskip 0.3cm
\noindent
Let $V$ and $G$ be two vector bundles over a compact Riemannian 
manifold $M$ with smooth boundary, such that
${\rm dim}\,V>{\rm dim}\,G$. Consider a bosonic gauge theory and let the 
first-order differential operator
$R:\ C^\infty(G,M)\to C^\infty(V,M)$ be the generator of infinitesimal
gauge transformations.
Let $\Delta:\ C^\infty(V,M)\to C^\infty(V,M)$ be the 
gauge-invariant second-order differential operator of the linearized
field equations. Let the
operators ${L} \equiv \bar R R:\ C^\infty(G,M)\to C^\infty(G,M)$
and $F \equiv \Delta+R\bar R$ be of Laplace type and normalized by 
$\sigma_L({L})=|\xi|^{2} I_{G}$.
Let $\sigma_L(R;N)\Big|_{\partial M}\equiv \nu$
and $\sigma_L(R;\zeta)\Big|_{\partial M}\equiv \mu$ be the restrictions
of the leading symbol of the operator $R$ to the boundary,
$N$ being the inward-pointing unit normal to the boundary
and $\zeta\in T^{*}(\partial M)$ being a cotangent vector, and 
$\Pi=I_V -\nu\bar\nu$.

Then the generalized boundary-value problem $(F,B_F)$ with the boundary
operator $B_F$ determined by the boundary 
conditions (48) and (51)
is gauge-invariant provided that the ghost boundary operator 
$B_{L}$ takes the Dirichlet form. 
Moreover, it is strongly elliptic with respect to the cone ${\bf C}
-{\bf R}_{+}$
if and only if the matrix $[|\zeta| I_V +i\nu\bar\nu\mu\bar\nu]$
is positive-definite.
A sufficient condition for that reads
$$
(I_V-\Pi)[|\zeta|^{2} I_V-\mu\bar\mu](I_V-\Pi)>0.
\eqno (63)
$$

In Euclidean quantum gravity, however, if a gauge-averaging
functional of the de Donder type is chosen, with a gauge parameter
such that the resulting operator on metric perturbations is of
Laplace type, the boundary-value problem with boundary conditions
(48) and (51) fails to be strongly elliptic [1]. Such a result
raises deep interpretative issues, since it seems to imply that
a Becchi--Rouet--Stora--Tyutin-invariant quantization of the
gravitational field cannot be implemented on manifolds with
boundary (cf. Ref. [3]). Two alternatives seem therefore to emerge:
\vskip 0.3cm
\noindent
(i) use local boundary conditions which are not completely
gauge-invariant because they do not involve tangential derivatives and
hence preserve strong ellipticity. The corresponding quantum
amplitudes should however be gauge-invariant [4].
\vskip 0.3cm
\noindent
(ii) study non-local boundary conditions [5] and try to understand 
whether they are compatible with ellipticity,
self-adjointness and with the
request of invariance under a suitable class of transformations.
\vskip 0.3cm
\noindent
It seems therefore that we are still at the very beginning in 
the process of understanding the interplay between the problems
of Euclidean quantum gravity on the one hand, and the problems
of global analysis on the other hand (different aspects of
the same problem have been discussed in Refs. [6,7]).
Hopefully, new perspectives in fundamental
physics and spectral geometry will be gained from such efforts.
\vskip 0.3cm
\leftline {\bf Acknowledgments}
\vskip 0.3cm
\noindent
The work of G. E. has been partially supported by PRIN97
`Sintesi'. Special thanks are due to Antonio Masiello for warm
hospitality in Monopoli, and to Emanuele Sorace for 
scientific advice.
\vskip 0.3cm
\leftline {\bf Appendix}
\vskip 0.3cm
\noindent
In the presence of a boundary for a Riemannian manifold $M$ of
dimension $m$, the local coordinates $x^\mu$, $(\mu=1,\dots,m)$,
are split into local coordinates $\hat x^k$, $(k=1,\dots,m-1)$, 
on $\partial M$, and the geodesic distance to the
boundary, $r$. Similarly, the coordinates $\xi_{\mu}$ on the cotangent
bundle $T^{*}(M)$ are split into coordinates $\zeta_{k}$ on 
$T^{*}({\partial M})$, jointly with a real parameter $\omega$.
The notion of ellipticity we are interested in
requires now that the leading symbol $\sigma_{L}(F)$ of an operator
$F$ of Laplace type should be elliptic in the interior of $M$, and
that a unique solution should exist of the ordinary differential equation:
$$
\left[-{d^{2}\over dr^{2}}
+|\zeta|^{2}-\lambda\right]\varphi(r)=0,
\eqno ({\rm A}.1)
$$
subject to the boundary conditions and to a decay condition
at infinity. A thorough formulation of boundary conditions needs
indeed some abstract thinking. For this purpose, one has to
consider two vector bundles $W_{F}$ and $W_{F}'$ over the 
boundary of $M$, with a {\it boundary operator} $B_{F}$ relating
their sections, i.e.
$$
B_{F}: C^{\infty}(W_{F},{\partial M}) \rightarrow
C^{\infty}(W_{F}',{\partial M}).
$$
All the information about normal derivatives of the fields is not
encoded in $B_{F}$ but in the {\it boundary data}
$\psi_{F}(\varphi) \in C^{\infty}(W_{F},{\partial M})$. In our
analysis one has
$$
\psi_{F}(\varphi)=\pmatrix{[\varphi]_{\partial M} \cr
[\varphi_{;N}]_{\partial M} \cr},
$$
and the strong ellipticity conditions demands that a unique
solution of Eq. (A.1) should exist, subject to the boundary
condition
$$
\sigma_{g}(B_{F})\psi_{F}(\varphi)
=\psi_{F}'(\varphi) \qquad
\forall \psi_{F}'(\varphi) \in 
C^{\infty}(W_{F}',{\partial M})
\eqno ({\rm A}.2)
$$
and to the asymptotic condition
$$
\lim_{r \to \infty} \varphi(r)=0.
\eqno ({\rm A}.3)
$$
With a standard notation, $\sigma_{g}(B_{F})$ is the 
{\it graded leading symbol} of the boundary operator $B_{F}$.
 When the boundary conditions
(53) are considered, one finds 
$$
\sigma_{g}(B_{F})=\pmatrix{ \Pi & 0 \cr
iT & I_V - \Pi \cr},
\eqno ({\rm A}.4)
$$
where $T$ is the anti-self-adjoint matrix defined in (57).

When all the above conditions are satisfied $\forall \zeta
\in T^{*}({\partial M}), \forall \lambda \in {\bf C}
-{\bf R}_{+}, \forall (\zeta,\lambda) \not = (0,0)$, the
boundary-value problem $(F,B_{F})$ is said to be strongly
elliptic with respect to the cone ${\bf C}-{\bf R}_{+}$.
This property is crucial to ensure the existence of the 
asymptotic expansion as $t \rightarrow 0^{+}$ of the $L^{2}$-trace,
${\rm Tr}_{L^2}\,\exp(-tF)$, which is frequently applied in quantum field
theory and spectral geometry.
\vskip 0.3cm
\leftline {\bf References}
\vskip 0.3cm
\noindent
\item {[1]}
Avramidi, I. G. and Esposito, G. (1999) {\it Commun. Math. Phys.}
{\bf 200}, 495.
\item {[2]}
Gilkey, P. B. and Smith, L. (1983) {\it J. Diff. Geom.}
{\bf 18}, 393.
\item {[3]}
Moss, I. G. and Silva, P. J. (1997) {\it Phys. Rev.} 
{\bf D 55}, 1072.
\item {[4]}
Luckock, H. C. (1991) {\it J. Math. Phys.} {\bf 32}, 1755.
\item {[5]}
Esposito, G. (1999) {\it Class. Quantum Grav.} {\bf 16}
(gr-qc/9806057).
\item {[6]}
Avramidi, I. G. and Esposito, G. (1998) 
{\it Heat-kernel asymptotics of the
Gilkey--Smith boundary-value problem}, to appear in: Proceedings
of the Conference ``Trends in Mathematical Physics'',
Knoxville, Oct. 14--17, 1998, 21 pp;  
Cambridge: International Press (1999) (math-ph/9812010).
\item  {[7]}
Avramidi, I. G. and Esposito, G. (1998) 
{\it On ellipticity and gauge invariance
in Euclidean quantum gravity}, to appear in: Proceedings
of the Conference ``Trends in Mathematical Physics'',
Knoxville, Oct. 14--17, 1998, 10 pp;
Cambridge: International Press (1999) (hep-th/9810009).

\bye